\documentclass[iop]{emulateapj}

\newcommand*{\teff}{$T_{\rm eff}$}
\newcommand*{\logg}{$\log~g$}
\newcommand*{\feh}{[Fe/H]}
\newcommand*{\kms}{km s$^{-1}$}
\newcommand*{\zmax}{$Z_{\rm max}$}
\newcommand*{\rmax}{$r_{\rm max}$}
\newcommand*{\rmin}{$r_{\rm min}$}

\newcommand*{\gaia}{$Gaia$}

\newcommand*{\nai}{\ion{Na}{1}}
\newcommand*{\nafe}{[Na/Fe]}

\newcommand*{\insitu}{$in~situ$}

\usepackage{natbib}
\usepackage{hyperref}
\usepackage{multirow}
\usepackage{tabularx}
\usepackage{color}

\hypersetup{
    bookmarks=true,         
    unicode=false,          
    pdftoolbar=true,        
    pdfmenubar=true,        
    pdffitwindow=true,     
    pdfstartview={FitH},    
    pdftitle={},    
    pdfauthor={Lee},     
    pdfsubject={Astronomy},   
    pdfcreator={dvipdf},   
    pdfproducer={dvipdf}, 
    pdfkeywords={},
    pdfnewwindow=true,      
    colorlinks=true,       
    linkcolor=red,          
    citecolor=blue,        
    filecolor=magenta,      
    urlcolor=cyan,           
    breaklinks=true,
    linktocpage
}

\shorttitle{Determination of Sodium Abundance Ratio} 
\shortauthors{Koo et al.} 

\begin{document}
\title{Determination of Sodium Abundance Ratio from Low-Resolution Stellar Spectra and Its Applications}
\author{Jae-Rim Koo\altaffilmark{1,2}, Young Sun Lee\altaffilmark{1}, Hye-Jin Park\altaffilmark{3}, Young Kwang Kim\altaffilmark{1}, and Timothy C. Beers\altaffilmark{4} \\
\scriptsize{
\textup{
\affil{1}{$^1$Department of Astronomy and Space Science, Chungnam National University, Daejeon 34134, Republic of Korea; youngsun@cnu.ac.kr} \\
\affil{2}{$^2$Department of Astronomy and Atmospheric Sciences, Kyungpook National University, Daegu 41566, Republic of Korea} \\
\affil{3}{$^3$Department of Physics and Astronomy, Sejong University, Seoul 05006, Republic of Korea} \\
\affil{4}{$^4$Department of Physics and JINA Center for the Evolution of the Elements, University of Notre Dame, IN 46556, USA}}}}


\begin{abstract} 

We present a method to determine sodium-abundance ratios ([Na/Fe]) using the \ion{Na}{1} D doublet lines 
in low-resolution ($R \sim$ 2000) stellar spectra. As stellar \nai\ D lines are blended with those produced 
by the interstellar medium (ISM), we developed a technique for removing the interstellar \nai\ D lines 
using the relationship between extinction, which is proportional to $E(B-V)$, and the equivalent width (EW) of 
the interstellar \nai\ D absorption lines. When measuring [Na/Fe], we also considered 
corrections for non-local thermodynamic equilibrium (NLTE) effects. 
Comparisons with data from high-resolution spectroscopic surveys suggest that the expected 
precision of [Na/Fe] from low-resolution spectra is better than 0.3 dex for stars with [Fe/H] $>$ --3.0. We also present a simple 
application employing the estimated [Na/Fe] values for a large number of stellar spectra from the Sloan Digital 
Sky Survey (SDSS). After classifying the SDSS stars into Na-normal, 
Na-high, and Na-extreme, we explore their relation to stars in Galactic globular clusters (GCs).  
We find that, while the Na-high SDSS stars exhibit a similar metallicity distribution function (MDF) 
to that of the GCs, indicating that the majority of such stars may have originated from 
GC debris, the MDF of the Na-normal SDSS stars follows that of typical disk and halo stars. As there 
is a high fraction of carbon-enhanced metal-poor stars among the Na-extreme stars, they
may have a non-GC origin, perhaps due to mass-transfer events from evolved binary companions.

\end{abstract}

\keywords{methods: data analysis --- technique: spectroscopy ---
galaxy: halo --- stars: abundances --- stars: kinematics and dynamics}

\section{Introduction}\label{sec1}
A non-zero fraction of field stars in the Milky Way (MW) 
are known to exhibit large abundances of light elements such as N, Na, and 
Al \citep{carretta2010,trincado2016,martell2016,schiavon2017a,pereira2019}.
Globular clusters (GCs) in the MW also display significant abundance 
variations of the light elements, as well as strong anti-correlations between the abundances 
of Na and O, N and C, and Al and Mg \citep{carretta2009}, which numerous authors have taken to 
be associated with the existence of multiple populations of (first- and second-generation) stars in GCs.

Because the second-generation stars 
in MW GCs exhibit relatively high abundances of N, Na, and Al, the 
field stars enhanced with such elements are often regarded 
as objects that have been accreted from tidally disrupted GCs 
\citep{carretta2009,carretta2016,martell2016,schiavon2017a,horta2021,tang2021}.
From this perspective, it is expected that the debris from partially or fully disrupted GCs contributed 
in some degree to the build-up of the Galactic halo \citep{thomas2020,wan2020} 
and bulge \citep{lee2019,lim2021}, in addition to stars accreted 
from major and minor merger events such as 
Gaia-Sausage \citep[GS;][]{belokurov2018}, Gaia-Enceladus \citep[GE;][]{helmi2018}, 
Sequioa \citep{myeong2019}, Thamnos 1 \& 2 \citep{koppelman2019}, and the currently disrupting 
Sagittarius dwarf galaxy \citep{ibata1994,majewski2003}.

Several previous studies have been carried out to evaluate the significance 
of the contribution of the GC-origin stars to the Galactic halo. For example, 
\citet{martell2010} searched for halo giants that originated from GCs by 
identifying N-rich stars on the basis of CN-band strengths in low-resolution stellar spectra from
the Sloan Digital Sky Survey (SDSS; \citealt{york2000}), in particular the stellar specific sub-survey 
Sloan Extension for Galactic Understanding and Exploration (SEGUE; \citealt{yanny2009}), and reported 
a fraction of 2.5\% of N-rich stars in the halo populations. \citet{horta2021} also attempted to 
estimate the contribution of the GCs to the halo using the nitrogen abundances of giant stars 
from  the Apache Point Observatory Galactic Evolution Experiment \citep[APOGEE;][]{majewski2017}, 
and derived a similar fraction, 2.7\%.

In addition to N, Na has become a 
key element for recognizing the population of stars that have escaped from MW GCs and to estimate 
the contribution of the GC-origin stars in the local halo, as it is 
relatively straightforward to measure its abundance in optical and near-IR spectra for dwarfs and giants. 
The Galactic field halo stars usually have sodium-abundance 
ratios ([Na/Fe]) in the range $-0.5 \lesssim {\rm[Na/Fe]} \lesssim +0.5$ \citep{suda2008}, while higher 
abundance ratios are observed among members of the Galactic GCs \citep{carretta2010}. Based 
on these results, several studies \citep{carretta2010, martell2011,ramirez2012} 
attempted to estimate the fraction of the Na-rich stars in the MW halo, reporting fractions 
from 2.8\% to 17\%. The clear interpretation is that there is a  non-negligible 
contribution of disrupted GCs to the assembly of the Galactic halo. However, 
since most of these studies are based on small numbers of relatively bright stars (with [Na/Fe] 
obtained from high-resolution spectroscopy), they are limited to relatively local samples. 
In order to obtain a better understanding of the connection of the GC debris to the assembly 
of the Galactic halo, it is necessary to study a much large number of homogeneously analyzed 
stars reaching farther into the Galactic halo.

Fortunately, spectroscopic surveys such as legacy SDSS, SEGUE, and the 
Large Sky Area Multi-Object Fiber Spectroscopic Telescope \citep[LAMOST;][]{cui2012} 
provide unprecedented numbers of stellar spectra,  with resolution sufficiently high  
to identify the \nai\ D resonance lines. However, in order to estimate [Na/Fe] from 
a low-resolution stellar spectrum, one challenge is the presence of sodium from the 
integrated interstellar medium (ISM) between the Earth and the stars. The \nai\ D lines 
produced in the stellar atmosphere are blended with those from Na in the ISM, and they are 
not easily distinguishable. As a result, the \nai\ D lines in a
low-resolution spectrum appear stronger than the intrinsic lines, and [Na/Fe] is generally 
over-estimated; the problem becomes even more severe at low Galactic latitudes. Note 
that throughout this paper, 
we refer to the \nai\ D lines produced by Na in the ISM as interstellar sodium (ISS) lines.

However, if one can remove the effects of ISS lines in a low-resolution spectrum, 
it is possible to obtain reasonably well-determined [Na/Fe] for 
very large numbers of stellar spectra from SDSS and LAMOST. In this 
paper, we collectively refer to all of the stellar spectra from SDSS, SEGUE, Baryon Oscillation 
Spectroscopic Survey \citep[BOSS;][]{dawson2013}, and extended Baryon Oscillation Spectroscopic 
Survey \citep[eBOSS;][]{blanton2017} as SDSS stars. Based on these 
large samples, one can better understand the distribution and properties of 
the Na-rich stars and their relationship to MW GCs, and derive 
statistically well-estimated fractions of these stars in the Galactic halo. This paper presents 
a method for removing the effect of ISS from low-resolution SDSS stellar spectra, 
and determine estimates of their [Na/Fe] abundance ratios. 

This paper is organized as follows. Section \ref{sec2} describes the method 
for the removal of the ISS lines from the SDSS spectra and the 
determination of [Na/Fe]. Validation of the estimated [Na/Fe] ratios, 
along with systematic and random errors, and the impact of the signal-to-noise 
ratio (SNR) of a stellar spectrum on the measured [Na/Fe], are presented in Section \ref{sec3}. 
In Section \ref{sec4}, we present one application of sodium-abundance ratios to understand their 
linkage with GC debris. A summary is given in Section \ref{sec5}.

\section{Methodology for Estimation of [Na/Fe]}\label{sec2}

\subsection{Impacts of Interstellar Sodium Absorption} \label{naremoval}

As stellar light travels through interstellar space, its intensity is 
attenuated by absorption or scattering by the ISM. As one result, in particular for  
certain chemical elements (e.g., Na and Ca), we expect ISM lines to be present in a stellar
spectrum. The strength of these lines is correlated with the amount of dust or gas along the 
line-of-sight. The absorption lines produced by the ISM are normally narrow and sharp due to the small internal 
velocity dispersion of gas in the ISM. Thus, we can identify and remove them from 
the spectrum of an astronomical object with relatively small differences
in the radial velocities between the object and the ISM gas clouds, especially for high-resolution spectra. 

However, at low resolution, such as for SDSS stellar spectra, the stellar absorption lines 
are blended with the ISM absorption lines. In this case, they are not easily distinguishable 
from the intrinsic stellar lines, even for a relatively high-velocity star. When measuring 
the abundance of a given chemical element, the overall effect of the unresolved ISM lines 
is to produce an over-estimate of the stellar 
elemental abundance. Sodium is a classic example of this difficulty in optical spectra.  

\begin{figure}[t!]
\includegraphics[width=\linewidth]{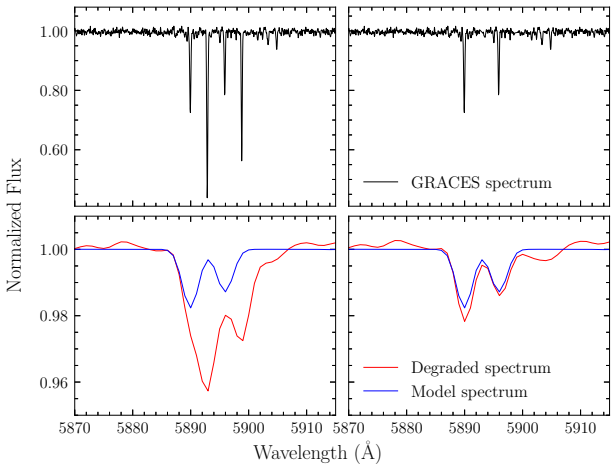}
\caption{Example spectra of the \nai\ D doublet region before (left column) and after (right column) 
removal of the ISS lines for the star BD$+$44$^\circ$ 493. The upper panels show a 
high-resolution ($R$\,=\,45,000) spectrum obtained with Gemini/GRACES, whereas the red 
spectra in the bottom panels show its degraded spectrum to $R$\,=\,2000. The blue line 
is the synthetic spectrum generated with $T_{\rm{eff}}$\,=\,5430 K, log\,$g$\,=\,3.4, [Fe/H]\,=\,--3.8, 
and [Na/Fe]\,=\,0.3, derived from the high-resolution spectroscopic 
analysis of this star by \citet{ito2013}.}
\label{bd44}
\end{figure}

The impact of the ISS lines on a stellar spectrum is illustrated in Figure \ref{bd44}. The top-left 
panel of Figure \ref{bd44} displays a high-resolution spectrum in the \ion{Na}{1} D doublet region 
of the extremely metal-poor star BD+44$^\circ$ 493, which was observed with 
Gemini/GRACES \citep{chene2014}. From inspection, one can 
easily identify the strong ISS lines shifted to the red by about 3 \AA\ from the 
weaker stellar lines. The red spectrum in the bottom-left panel shows a degraded version (to the SDSS 
resolving power of about $R\sim 2000$) of the high-resolution spectrum; the stellar 
and the ISS lines are indistinguishable.
The top-right panel of the figure exhibits the high-resolution spectrum after excising the 
ISS lines; the bottom-right panel shows the degraded spectrum (red line).
We clearly see the much weaker strength of the intrinsic lines compared to the ISS lines. 
The blue spectrum in the panels is a synthetic one generated with the stellar 
parameters ($T_{\rm{eff}}$\,=\,5430 K, log\,$g$\,=\,3.4, 
and [Fe/H]\,=\,--3.8) and sodium-abundance ratio ([Na/Fe]\,=\,+0.3), reported by 
\citet{ito2013}. Comparison of the bottom-left and bottom-right panels makes it obvious 
that the sodium-abundance ratio determined from low-resolution stellar spectra would result in 
a gross over-estimation without accounting for the presence of the ISS lines.

\begin{figure*}
\centering
\includegraphics[width=0.8\linewidth]{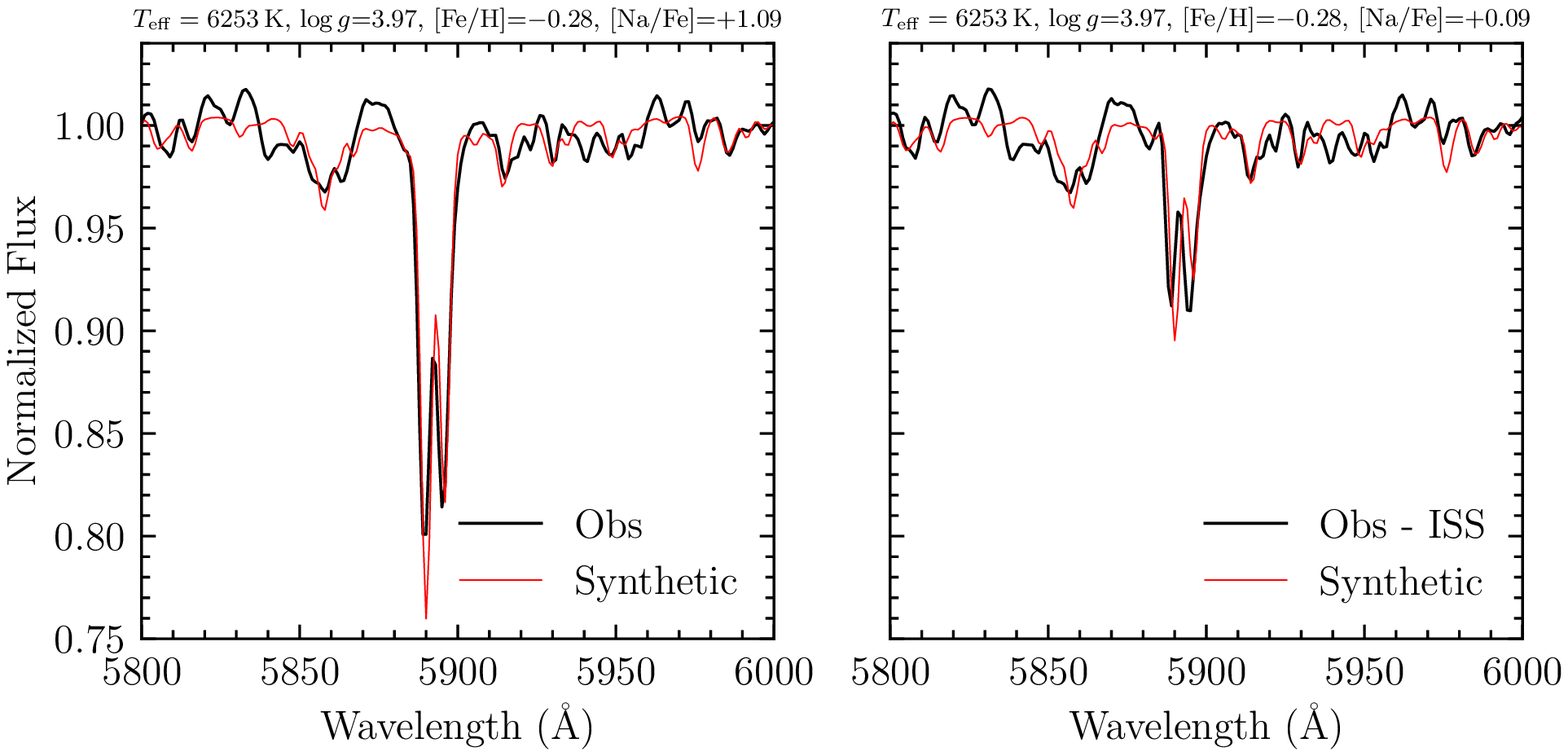}
\caption{An example of our ISS line-removal method applied to the spectrum of an SDSS star (J185226.5$+$201240). 
The left panel shows the spectrum before the removal of the ISS lines, while the right panel illustrates 
the spectrum after their removal. The black line indicates the observed spectrum; the red lines are the 
best-matching synthetic spectrum. The stellar parameters from the SSPP and [Na/Fe] (derived from the 
best-matching synthetic spectrum) are listed at the top of each panel. 
The much higher sodium-abundance ratio (by 1.0 dex) for the spectrum without removal of the 
the ISS lines is evident.}
\label{ex_sdss}
\end{figure*}


\subsection{Removal of Interstellar Sodium Absorption} \label{rev_na}

As demonstrated above, the removal of the \nai\ D lines produced by ISM is straightforward 
for high-resolution spectra, since we can easily identify the lines and excise them. In the 
case of low-resolution spectra, the situation is more complicated because the ISS 
lines are unresolved. Therefore, we require a method for eliminating the ISS lines 
before the estimation of \nafe, and have developed the following approach. 

We first estimated
the strength of the ISS lines in terms of the expected equivalent width (EW) along the line-of-sight to 
a given star, and subtracted it from the measured \nai\ D lines present in a stellar spectrum.
In order to calculate the EW of the ISS lines at a given Galactic coordinate, 
we adopted from the literature the relations between the reddening, $E(B-V)$, and the 
strength of the \nai\ D absorption lines by the ISM. \citet{poznanski2012} derived a 
correlation between the reddening estimated by \citet{sfd1998} and the EW of the 
ISS lines, using high-resolution spectra of quasars (QSOs) and 
low-resolution spectra of QSOs and galaxies obtained by SDSS. \citet{murga2015} also 
presented a relation between the reddening and the strength of the interstellar 
absorption of the \nai\ D and \ion{Ca}{2} H \& K lines, which were derived using extra-galactic sources from 
SDSS. They provide a best-fit relation for the range of $ 0 \le E(B-V) < 0.08$. 
We employed the results from both studies. For stars with  $ 0 \le E(B-V) < 0.08$
we adopted the relation of \citet{murga2015}:

\begin{eqnarray*} 
 \mathrm{EW}(E(B-V))_{\rm{D2}} = \left(\frac{E(B-V)}{0.39}\right)^{0.63}\, \mathrm{\AA}\\
 \mathrm{EW}(E(B-V))_{\rm{D1}} = \left(\frac{E(B-V)}{0.26}\right)^{1.06}\, \mathrm{\AA}\\
\end{eqnarray*}

\noindent For stars with $E(B-V)\geq0.08$ we adopt the relation of \citet{poznanski2012}:

\begin{eqnarray*} 
 \mathrm{EW}(E(B-V))_{\rm{D2}} = \frac{\mathrm{log_{10}}(E(B-V))+1.91}{2.16}\, \mathrm{\AA}\\
 \mathrm{EW}(E(B-V))_{\rm{D1}} = \frac{\mathrm{log_{10}}(E(B-V))+1.76}{2.47}\, \mathrm{\AA}.\\
\end{eqnarray*}

\noindent The subscript D1 in the equations indicates the \nai\ D line at 5896 \AA, while D2 applies for the line at 5890 \AA.

\begin{figure}
\centering
\includegraphics[width=0.95\linewidth]{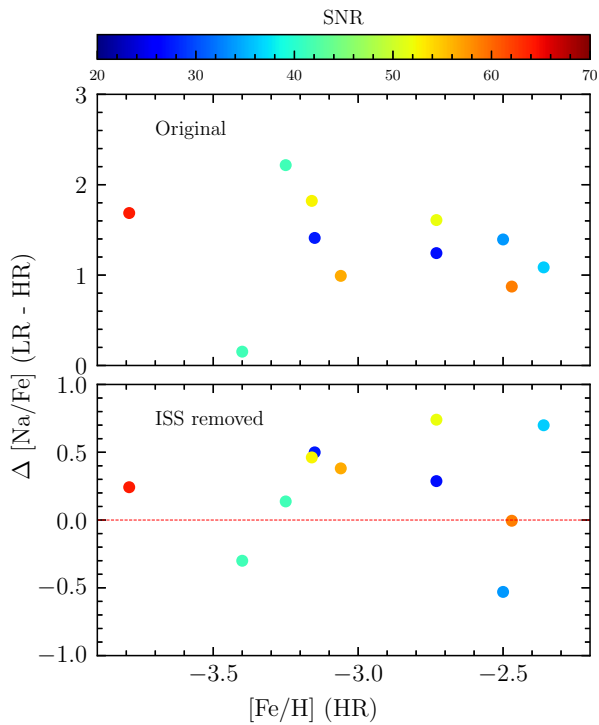}
\caption{Comparison of [Na/Fe] determined from the high-resolution (HR) analysis 
to that from our analysis of the degraded low-resolution (LR) spectrum before (upper panel) and after (lower panel) 
the ISS line removal. To remove the ISS lines, 
we first measured their EWs in the original high-resolution spectrum, and 
derived the Gaussian line profiles by following the method described in Section \ref{rev_na}. 
The derived line profiles were degraded and subtracted from the degraded high-resolution spectrum 
to obtain a LR spectrum with the ISS lines removed.
The color bar at the top represents the SNR of the high-resolution spectrum. 
It is clear that the difference becomes much smaller after the ISS lines are 
removed. There are no apparent trends with SNR.}
\label{ismcut}
\end{figure}


To remove the unresolved ISS lines in the low-resolution SDSS stellar spectra, 
from the above equations we first estimated the EW of the \nai\ D doublet lines 
produced by the ISM after computing the $E(B-V)$ value at a star's Galactic coordinates.
Because the $E(B-V)$ value derived from \citet{sfd1998} is the total reddening in a given direction, 
but the stars have different distances, we used instead the 3D reddening map from \citet{green2019} 
to determine the $E(B-V)$ value for a star using the python package DUSTMAPS \citep{green2018}.
The distance of a given star primarily comes from the parallax provided by $Gaia$ Early Data Release 3 \citep[EDR3;][]{gaia2021}. 

After obtaining the EW of the \nai\ D lines caused by the ISM along the line-of-sight to 
a star, we generated very narrow artificial \nai\ D doublet lines by forcing two Gaussian 
functions to have the same area as the calculated EW. We then degraded 
the Gaussian profiles to the SDSS resolution of $R$ = 2000, and subtracted it from the 
observed \nai\ D lines in a normalized SDSS spectrum to create an approximately ISS-free spectrum. 
While fitting the two Gaussians, we set the full width at half maximum (FWHM) to 
0.1 \AA. It turned out that the choice of the FWHM parameter is not critical, as we degrade the 
artificially created lines to a much lower resolution. In addition, 
we simply ignored the velocity of ISM, even though it can cause the ISS lines 
to be shifted, not only because it is infeasible to derive the radial velocities of the gas clouds 
in the ISM that contribute to the ISS lines, but it is known that the mean shift of the ISS lines in the Northern 
Hemisphere is about --3 \kms\ \citep{murga2015}, which is a very small offset in 
low-resolution spectra. 

Figure \ref{ex_sdss} shows an example of the removed ISS lines 
from a relatively metal-rich SDSS stellar spectrum (SDSS J185226.49$+$201240.2), following the method described above.  
The left panel of the figure shows the observed spectrum with the ISS lines included, while the right panel 
is the same spectrum after their removal. The effect of the removal is clear, with a substantially 
weaker Na doublet visible in the right panel.  The red line
in each panel is the best-fit model spectrum. The stellar atmospheric 
parameters ($T_{\rm{eff}}$, log\,$g$, and [Fe/H]) and the sodium-abundance ratio used to 
generate the synthetic spectrum are listed at the top 
of each panel. The stellar atmospheric parameters  for the 
SDSS spectrum were determined using the latest version of the SEGUE  
Stellar Parameter Pipeline \citep[SSPP;][]{allende2008,lee2008a,lee2008b,lee2011,smolinski2011}. 
The sodium-abundance ratio was estimated by following the method described in Section \ref{na_det}. 
It is obvious that a much higher estimate of \nafe\ is obtained for the spectrum including the 
ISS lines, as expected. 

\begin{figure*}[!t]
\centering
\includegraphics[width=0.9\linewidth]{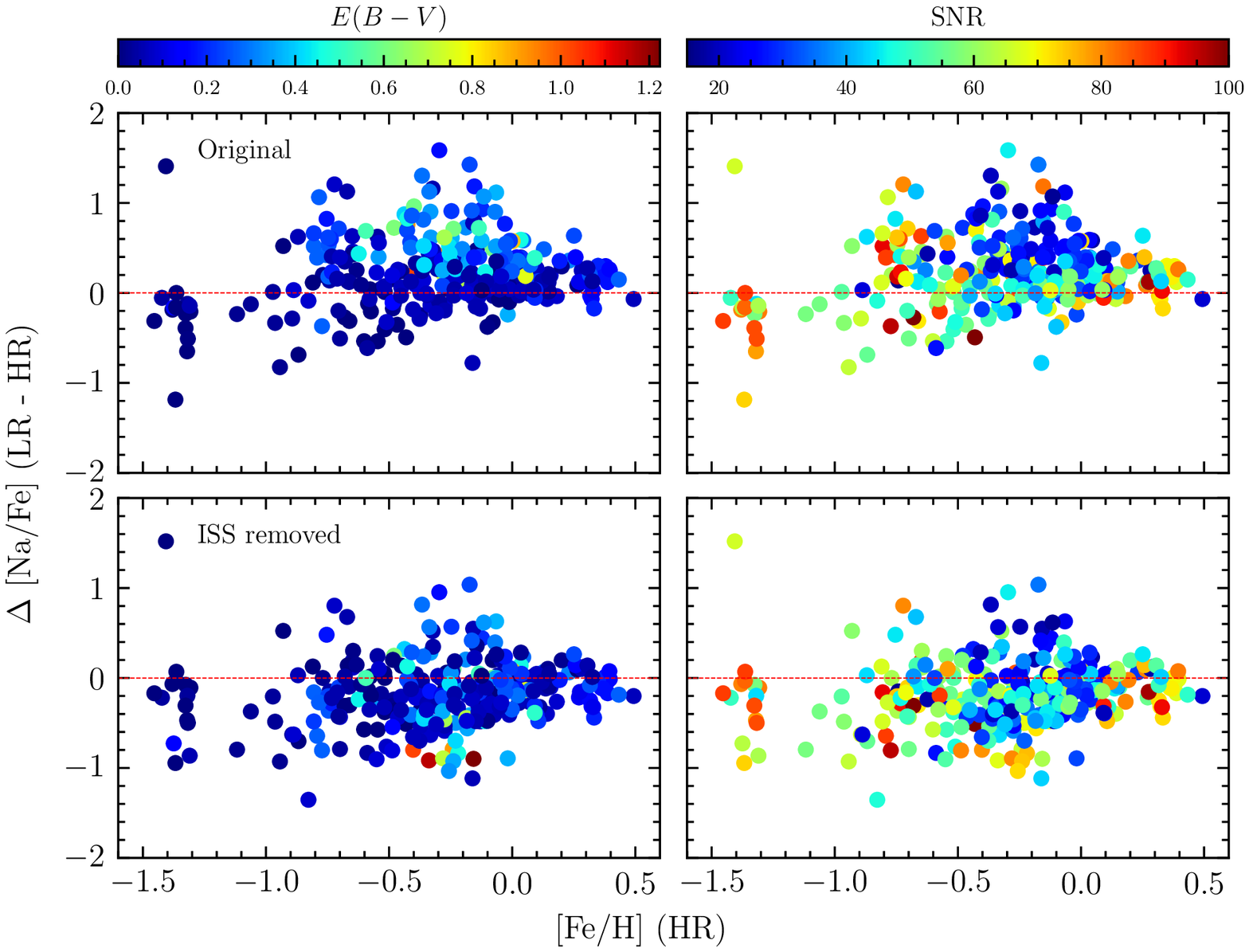}
\caption{Same as in Figure \ref{ismcut}, but for comparison with APOGEE. The color bar at the 
top left is added and represents the 3D reddening value derived from \citet{green2019} after adopting
\gaia\ parallaxes. There are 299 stars included.}
\label{nafeapo}
\end{figure*}


\subsection{Method for Determination of [Na/Fe]} \label{na_det}

In order to estimate [Na/Fe] from SDSS stellar spectra with the ISS lines 
removed, we applied a similar spectral-matching technique as that currently used in the SSPP \citep{lee2011,lee2013}.
Briefly, we first generated 
a grid of synthetic spectra with wide ranges of stellar parameters and [Na/Fe] values. We employed 
the Kurucz NEWODF model atmospheres \citep{castelli2003}, which assumes 
plane-parallel line-blanketed model structures in one-dimensional local thermodynamical 
equilibrium (LTE). Then, we utilized the TURBOSPECTRUM synthesis code \citep{alvarez1998}, 
which adopts the solar abundances of \citet{asplund2005}, to 
create a synthetic spectrum in the wavelength range between 5500 \AA\ and 6500 \AA\ 
at a resolving power of $R$ = 500,000. The generated grid covers 4000 K $\leq T_{\rm{eff}} \leq$ 7000 K in 
steps of 250 K, $0.0 \leq$ log\,$g$ $\leq 5.0$ in steps of 0.5 dex, --5.0 $\leq$ [Fe/H] $\leq$ +1.0 in steps 
of 0.5 dex, and --1.0 $\leq$ [Na/Fe] $\leq$ +3.0 in steps of 0.5 dex. A total 
of 16,731 synthetic spectra were generated. Then, these spectra were degraded to the SDSS 
resolution ($R$ = 2000) and normalized with a pseudo continuum, which was derived 
from iterative polynomial fitting to a spectrum. 

After the radial-velocity correction is made for an observed spectrum, we applied the 
same continuum routine used for the synthetic grid to the observed SDSS spectrum 
to obtain a normalized spectrum. From the normalized spectrum, we 
removed the ISS lines following the methodology described in Section \ref{rev_na}.
The best-matching model spectrum was searched for by minimizing 
the residuals in the wavelength range of 5880 \AA\ -- 5910 \AA\ between the 
normalized observed spectrum and the synthetic template, using the IDL/MPFIT routine \citep{markwardt2009}. 
The spectral range considered is sufficiently wide to define an accurate 
continuum and include the \nai\ D lines. During the minimization, we held $T_{\rm{eff}}$, log\,$g$, 
and [Fe/H] constant, and only varied [Na/Fe] to search for the best-fit 
model. The stellar parameters come from the latest version of the SSPP.


\begin{table*}
\centering
\caption{Systematic and Random Errors of [Na/Fe] due to Errors in Stellar Parameters}
\label{tab1}
\begin{tabularx}{0.78\textwidth}{c>{\hspace{1pc}}ccc>{\hspace{1pc}}ccc>{\hspace{1pc}}ccc} 
\hline
{}     & {$T_{\rm{eff}}$} & \multicolumn{2}{c}{$\Delta$[Na/Fe]}& {log\,$g$}& \multicolumn{2}{c}{$\Delta$[Na/Fe]} & {[Fe/H]} & \multicolumn{2}{c}{$\Delta$[Na/Fe]} \\ 
\cline{3-4} \cline{6-7} \cline{9-10}
{}     & {Error} & {$\mu$} & {$\sigma$} & {Error} & {$\mu$} & {$\sigma$} & {Error} & {$\mu$} & {$\sigma$} \\
{Type} & {(K)} & {(dex)} & {(dex)} & {(dex)} & {(dex)} & {(dex)}& {(dex)} & {(dex)} & {(dex)}  \\
\hline
{Giant}      &   {+100} &   {+0.161} & {0.035}   &   {+0.2} & {$-$0.023}   & {0.047} & {+0.2}   & {$-$0.210}   & {0.037} \\
{Cool dwarf} &   {+100} &   {+0.155} & {0.035}   &   {+0.2} & {$-$0.055}   & {0.093} & {+0.2}   & {$-$0.182}   & {0.041} \\
{Hot dwarf}  &   {+100} &   {+0.106} & {0.019}   &   {+0.2} & {$-$0.083}   & {0.119} & {+0.2}   & {$-$0.190}   & {0.019} \\
{Giant}      & {$-$100} & {$-$0.164} & {0.039}   & {$-$0.2} &   {+0.017}   & {0.023} & {$-$0.2} &   {+0.201}   & {0.041} \\
{Cool dwarf} & {$-$100} & {$-$0.156} & {0.040}   & {$-$0.2} &   {+0.093}   & {0.047} & {$-$0.2} &   {+0.183}   & {0.041} \\
{Hot dwarf}  & {$-$100} & {$-$0.107} & {0.016}   & {$-$0.2} &   {+0.097}   & {0.023} & {$-$0.2} &   {+0.188}   & {0.024} \\
\hline
\end{tabularx}
\tablecomments{$\mu$ and $\sigma$ denote the mean and standard deviation from a Gaussian fit.}
\end{table*}


\subsection{Evaluation of the Interstellar Sodium Line-removal Scheme with High-resolution Spectra}
\subsubsection{Comparison with [Na/Fe] Derived from Gemini/GRACE Stellar Spectra}

As part of a study of the early chemical evolution of the MW, we have obtained 
high-resolution Gemini/GRACES spectra for several dozen extremely 
metal-poor (EMP; [Fe/H] $<$ --3.0) candidates selected from SDSS and LAMOST, and 
analyzed these spectra with MOOG \citep{sneden1973} to derive stellar 
parameters and abundances for various chemical elements, including Na (M. Jeong et al. in preparation). 
Eleven  of the EMP candidates have [Na/Fe] $>$ --0.5, and we make use of them 
to evaluate the performance of our ISS line-removal scheme by 
comparing our [Na/Fe] estimates from the low-resolution spectra before and after removing 
the ISS lines with those of the high-resolution analysis.

To remove the ISS lines from the Gemini/GRACES spectra in a similar fashion as we do for observed 
SDSS stars,  we first measured their EWs in the original high-resolution spectrum, 
and derived the two Gaussian line profiles following the method described in 
Section \ref{rev_na}. These derived line profiles were convolved to the 
SDSS resolution and subtracted from the degraded high-resolution spectrum. 
Following this step, we determined [Na/Fe] via our spectral-fitting technique. While 
determining [Na/Fe], we fixed the stellar parameters adopted from the high-resolution 
analysis. We also determined \nafe\ from the degraded high-resolution spectrum 
including the ISS lines.

Figure \ref{ismcut} shows the results of this exercise. 
The color bar shown above the top panel represents the SNRs of the high-resolution spectra. 
The upper panel shows the difference in the derived \nafe\ between our low-resolution 
estimate and that of the high-resolution analysis before the ISS line removal. 
It is apparent that the ISS lines overwhelm the intrinsic stellar \nai\ D lines, 
resulting in an estimated \nafe\ that is systematically higher compared to 
the high-resolution spectra, by over 1 dex. In the lower panel, which applies to the situation after 
the removal of the ISS lines, there is a much smaller offset (less than 
0.5 dex for most of the stars). Although we did not find any trend with SNR, our estimate of \nafe\ 
is slightly larger, on average, by about 0.2 dex.

We also found that the EWs computed from the ISS lines in the high-resolution spectra 
are actually quite close to those predicted from the relations between the reddening 
and EWs of the ISS lines described above. The sum of EWs of the doublets calculated 
from the 3D extinction map is on average about 6\% larger than that 
measured for the high-resolution spectrum. This result 
shows that our method for ISS line removal works quite well.  

\subsubsection{Comparison with [Na/Fe] Derived from APOGEE}

We also carried out a similar set of evaluations using data from APOGEE. The APOGEE 
survey obtains H-band spectra with a resolving power of $R \sim$ 22,500. Even though the APOGEE Data 
Release 16 \citep[DR16;][]{ahumada2020} lists [Na/Fe] abundances determined from different Na 
lines (16378 \AA\ and 16393 \AA), this test still provides a valuable comparison.  
Note that the Na abundance we adopted from APOGEE is based on the LTE assumption. 

Figure \ref{nafeapo} shows the difference between the APOGEE results and our measured \nafe\ 
from the \nai\ D lines with (top panels) and without (bottom panels) the ISS lines, similar 
to Figure \ref{ismcut}. From inspection, it is obvious that our estimated \nafe\ is consistently 
higher in the top panels, in particular for the stars with the higher reddening, as expected. We also note 
from inspection of the bottom-left panel that our ISS removal scheme has a tendency to slightly 
over-estimate the EW of the ISS for high-reddening stars, as they exhibit lower \nafe, on average. 
This behavior is opposite to that seen in Figure \ref{ismcut}, where 
we removed the predicted ISS lines from the high-resolution spectrum. For our 
present investigation (comparison of field stars with MW GCs, as described below), we are more interested in 
stars with [Fe/H] $>$ --2.5; thus we rely on the comparison with 
the APOGEE for further calibration (see Section \ref{hrcheck}). We note that no significant trends 
with [Fe/H] or SNR are apparent in the right-column panels.


\begin{table}
\centering
\caption{Impacts of SNR for Low-Resolution Spectra on Determination of [Na/Fe]}
\label{tab2}
\begin{tabular}{ccccccc} 
\hline
{}    & \multicolumn{6}{c}{$\Delta$[Na/Fe]} \\
{}    & \multicolumn{2}{c}{Giant} & \multicolumn{2}{c}{Cool dwarf} & \multicolumn{2}{c}{Hot dwarf} \\
\cline{2-3} \cline{4-5} \cline{6-7}
SNR   & {$\mu$} & {$\sigma$}& {$\mu$} & {$\sigma$}& {$\mu$} & {$\sigma$} \\
\hline
10 &  ~~0.022 &   0.036 &  ~~0.039 & 0.049 &  ~~0.076 & 0.075 \\
15 &  ~~0.006 &   0.029 &  ~~0.013 & 0.036 &  ~~0.034 & 0.058 \\
20 & $-$0.002 &   0.025 &  ~~0.002 & 0.030 &  ~~0.017 & 0.054 \\
30 & $-$0.009 &   0.020 & $-$0.003 & 0.026 & $-$0.000 & 0.038 \\
40 & $-$0.011 &   0.017 & $-$0.006 & 0.020 & $-$0.006 & 0.031 \\
50 & $-$0.012 &   0.017 & $-$0.005 & 0.019 & $-$0.010 & 0.027 \\
\hline 
\end{tabular}
\tablecomments{$\mu$ and $\sigma$ denote the mean and standard deviation from a Gaussian fit.}

\end{table}

\section{Validation of [Na/Fe] Determinations from Low-Resolution Spectra}\label{sec3}

\subsection{Random and Systematic Errors on [Na/Fe]}\label{syserror}

The strength of the \nai\ D lines increases at higher [Fe/H] and [Na/Fe], 
as expected, and are weaker for giants compared to dwarfs. 
Thus, a degeneracy could exist in the strength 
of the \nai\ D lines depending on stellar parameters, which could result in a systematic error 
in the estimated [Na/Fe] by our spectral-fitting method. In addition,  
as we fixed \teff, \logg, and \feh\ while estimating \nafe, the existence of any 
systematic errors in \teff, \logg, and \feh\ could produce a systematic error in 
the determined \nafe. Therefore, we need to check how the systematic errors 
in the adopted stellar parameters and the possible degeneracy of the 
\nai\ D line strengths affects the estimated [Na/Fe]. 

To carry out this test, we first processed the synthetic spectra through our methods to obtain
an estimate of the (known) [Na/Fe] under various conditions. While estimating [Na/Fe], we shifted the model 
parameters by $\pm$100 K for $T_{\rm{eff}}$ and $\pm$0.2 dex for log\,$g$ 
and [Fe/H], respectively, in order to estimate how much the derived \nafe\ 
can be affected by errors of the adopted stellar parameters.

After completing the estimates of [Na/Fe] to check the collective 
behavior of the determined \nafe, we divided the measured \nafe\ values 
into three groups: giants for \teff\ $<$ 5500 K and \logg\ $<$ 3.5, cool dwarfs for \teff\ $<$ 5500 K and 
\logg\ $\geq$ 3.5, and hot dwarfs for \teff\ $\geq$ 5500 K and \logg\ $\geq$ 3.5. 
We did not include the spectra for models with [Fe/H] $<$ --3.0, because 
not only are such objects rare, but we are here primarily interested in the escaped stars from 
the Galactic GCs or the debris from the already-disrupted GCs, which 
presumably have [Fe/H] $>$ --3.0. 

Table \ref{tab1} summarizes the results of the above exercise. 
It generally indicates that the systematic offsets of 
the measured \nafe\ do no vary much among the different groups, and that the random 
scatters are also small for all groups. Specifically, the incorrect 
assignment of surface gravity does not much affect the measured [Na/Fe], as the 
mean offset is less than 0.1 dex. The errors caused by the \teff\ and 
[Fe/H] offsets become a little larger, up to $\pm$0.16 dex and $\pm$0.2 dex, respectively, but 
are still not significant. The mean $\mu$ and standard deviation $\sigma$ in the 
table are determined from a Gaussian fit to the differences between our 
estimates and the model values. Considering that the typical uncertainty of the elemental 
abundance from the low-resolution spectroscopy is about 0.3 dex, we
conclude that the systematic errors in the 
stellar parameters do not significantly affect the estimated [Na/Fe].

\begin{figure*}
\centering
\includegraphics[width=0.8\linewidth]{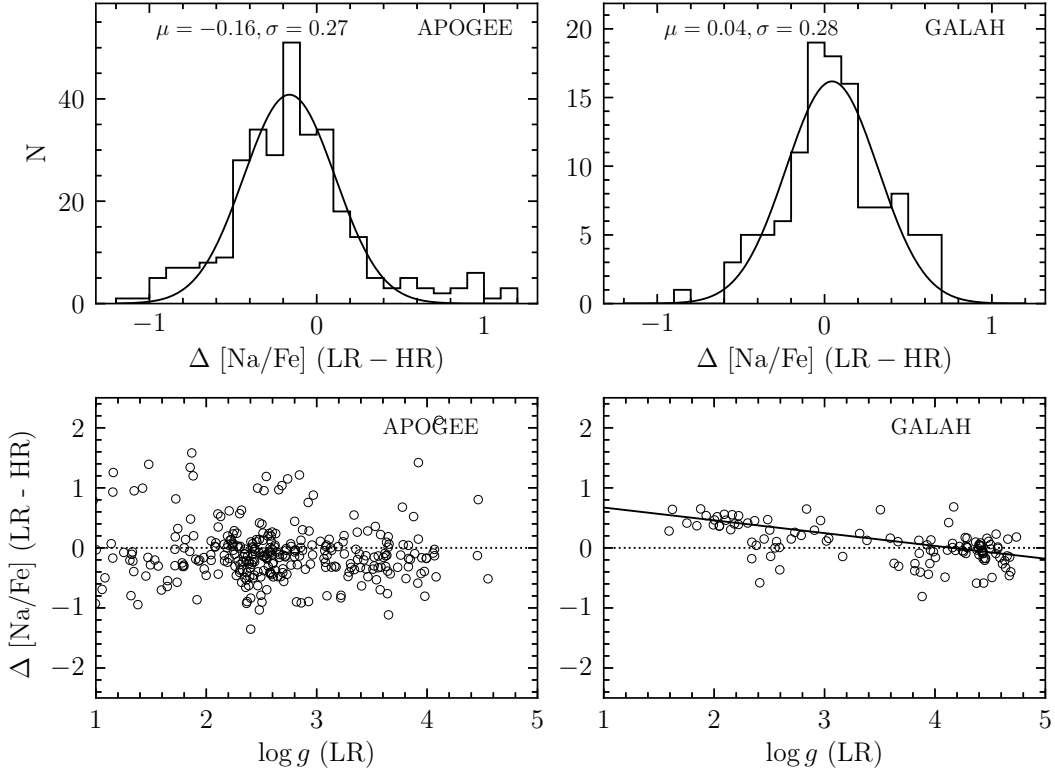}
\caption{Comparison of our derived sodium-abundance ratios with those from the APOGEE (left panels) 
and GALAH (right panels) surveys, which are based on high-resolution near-IR and optical 
spectroscopy, respectively. In the 
upper panels, the mean offset ($\mu$) and the 
standard deviation ($\sigma$) were obtained by fitting a Gaussian to both samples. 
The comparison with the APOGEE data exhibit no significant trend with \logg, so 
we decided to adjust the offset (--0.16 dex) to the [Na/Fe] value before applying 
the NLTE corrections described in the text (see Section \ref{nlte}. Since [Na/Fe] from the 
GALAH survey is based on NLTE, we compared our values of [Na/Fe] after the NLTE correction 
to those from the GALAH stars. The top-right 
panel indicates a small offset (less than 0.1 dex). However, we note a negative trend 
with surface gravity in the bottom-right panel. We fitted a linear function to the 
residuals to obtain a slope of --0.21 dex dex$^{-1}$ and intercept of 0.88 dex. We applied this 
correction function to our NLTE corrected [Na/Fe], and use this value for further analysis.}
\label{nafehr}
\end{figure*}


\begin{figure*}
\centering
\includegraphics[width=0.9\linewidth]{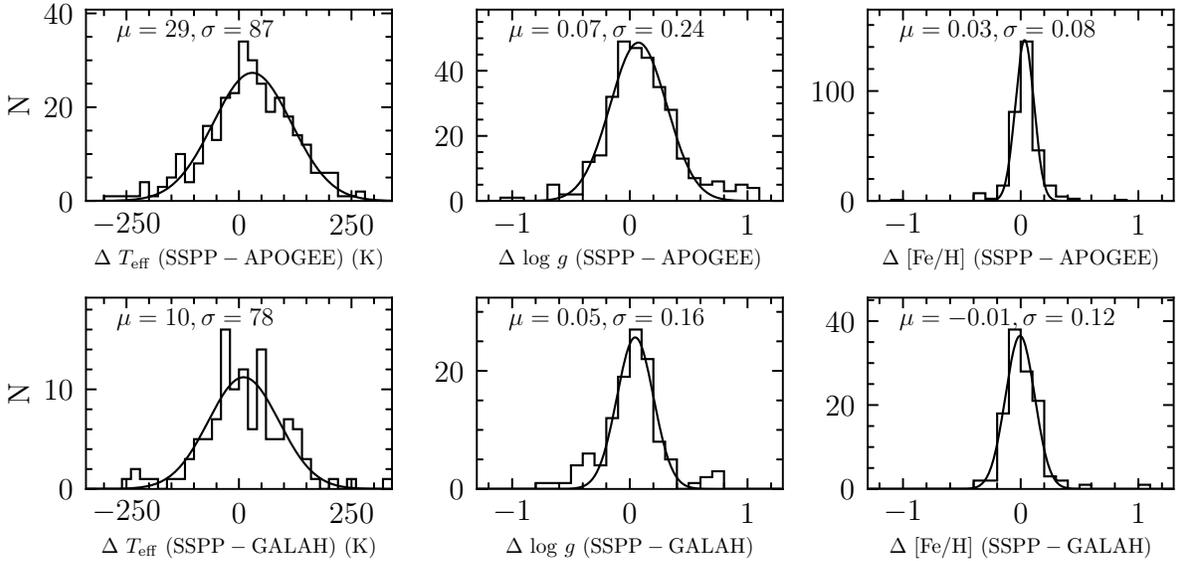}
\caption{Comparison of the stellar parameters from the SSPP with those from the APOGEE (upper panels) 
and GALAH (lower panels) data based on high-resolution spectroscopic analysis. 
The mean offset and the scatter for each parameter are derived from a Gaussian fit. Overall, 
we observe relatively small offsets and scatters from both samples for all three parameters. 
This figure only includes the stars with [Na/Fe] available in 
the APOGEE and GALAH data sets, as we calibrate our estimates of 
\nafe\ with those from these two surveys. Note that we used the NLTE [Fe/H] values 
from the GALAH survey; see the text for details on the validity of its use.}
\label{tgm}
\end{figure*}

\subsection{Impact of SNR on Estimated [Na/Fe]}\label{snrtest}

Stellar spectra from large sky surveys such as SDSS and LAMOST cover
a wide range of SNRs. Because the noise in a spectrum is able to induce systematic and random errors, 
we also check on how the SNR affects the estimated [Na/Fe]. 

To evaluate the impact of the SNR on the abundance estimate for sodium, 
following \citet{lee2008a}, we injected different levels of random noise into the grid of 
synthetic spectra used for the spectral fitting. In this process, we generated a set of 
10 different noise-added synthetic spectra 
for SNR = 10, 15, 20, 30, 40, and 50 per given model spectrum. 
Then, we applied our method to the noise-added synthetic spectra to recover [Na/Fe]. 
In our analysis, we took an average value of 10 [Na/Fe] values from 10 simulated 
noise-injected spectra at a given SNR. To analyze the results of this test, 
once again we divided the test sample into three groups 
as in Section \ref{syserror}: giants, cool dwarfs, and hot dwarfs. 

Table \ref{tab2} summarizes the results of the above exercise.  It indicates that, even though 
there is a tendency that the mean offset and scatter increase with decreasing SNR, 
as expected, their size is relatively small. The mean $\mu$ and standard deviation $\sigma$ in the 
table are determined from a Gaussian fit to the residuals between our estimates and the model 
parameters for [Fe/H]$\geq$ --3.0. Furthermore, we observe that, as the metallicity decrease and 
temperature increases, the mean difference and their scatter become larger, presumably due to 
the weaker \nai\ D lines. Nonetheless, the mean offset and scatter are significantly less than 0.1 dex 
for all three groups and SNRs. Consequently, this exercise confirms that our method is 
very robust to the SNR of the spectrum. 

\subsection{Non-LTE Corrections}\label{nlte}

It is known that the [Na/Fe] abundance ratio can vary, by as much as 0.5 dex, under the non-LTE (NLTE) assumption, 
especially when the EW of the \nai\ D lines is around 150 to 200 m\AA\,\citep{gratton1999, lind2011}. 
The correction is generally larger for giants than dwarfs \citep{amarsi2020}. Therefore, 
we need to apply NLTE corrections to our measured \nafe. 
\citet{gratton1999} reported the NLTE corrections for lines of 
\ion{Fe}{1}, \ion{O}{1}, \ion{Na}{1}, and \ion{Mg}{1}, and provided the correction values 
for stars with \teff\ from 4000 K to 7000 K with 1000 K steps, 
over a broad range of gravities (\logg\ = 1.5, 3.0, 4.5) and metallicities ([Fe/H] = 0, --1, --2, --3). 
For the \nai\ D lines the correction grid is valid for the EW range of 100 m\AA\ -- 1350 m\AA . 
We adopted this grid to derive the NLTE correction value by linear 
interpolation over the stellar parameters and the EW of the \nai\ D lines.
When obtaining the correction value from this interpolation, we used 
the EW of the \nai\ D lines calculated from the synthetic spectrum, which is generated with the 
estimated \teff, \logg, \feh, and \nafe\ for each star, because 
it provides much clearer lines to calculate the EW. We simply took the value 
from the grid that is the nearest to \teff, \logg, [Na/Fe], and EW for 
the star whose parameters are outside the grid.

\subsection{Calibration of Estimated [Na/Fe] with High-resolution Surveys} \label{hrcheck}

Comparison of our derived [Na/Fe] with that estimated based on high-resolution 
spectroscopy provides a direct measure of the reliability of our estimated \nafe.  We 
employed stars observed in two surveys, APOGEE and the Galactic Archaeology 
with HERMES \footnote{\url{https://galah-survey.org/}} \citep[GALAH;][]{zucker2012} survey.
The GALAH survey obtained optical spectra with a resolving power of $R \sim$ 28,000. 
Although the APOGEE DR16 and the GALAH Data Release 3 \citep{buder2021} 
contain [Na/Fe] abundances determined from different Na lines: 16378 \AA\ and 16393 \AA\ 
for APOGEE and 5688 \AA, 6154 \AA, and 6169 \AA\ for GALAH, we expect the derived Na abundance 
ratio should be the same independent of the lines employed, unless some unexpected problem exists. 
Below we describe a comparison exercise between the two surveys. 

Cross matches (within 1$\arcsec$) were used to identify 318 and 116 stars 
in common between SDSS and APOGEE, and SDSS and GALAH, respectively. We selected only the stars 
with \teff\ = 4000 -- 7000 K and the uncertainties of the measured \nafe\ less than 0.1 dex 
in the comparison-survey data. These stars also met the requirement of 10 \kms\ in radial-velocity 
differences. For the SDSS spectra, we removed the ISS lines following the scheme 
described in Section \ref{rev_na}. We adopted for each star the inverse parallax from 
$Gaia$ EDR3 as the distance used to calculate the strength of the ISS 
lines. The sodium-abundance ratio was then determined by fixing the atmospheric 
parameters (\teff, \logg, and \feh) obtained from the SSPP. 

Note that the [Na/Fe] abundance ratio from APOGEE is based on an LTE assumption, while that 
from the GALAH is based on the NLTE assumption. To carry out the calibration properly, we first 
compared our measured value before the NLTE correction with that of APOGEE to check if any systematic error exists in the 
LTE value. The left column of panels in Figure \ref{nafehr} show histograms of the \nafe\ 
differences (top panel) between our low-resolution (LR) and the APOGEE high-resolution (HR) 
value, and their trend with \logg\ (bottom panel). 
We found a systematic offset of --0.16 dex and a standard deviation of 0.27 dex, which 
were calculated with a Gaussian fit to the residuals. The bottom panel suggests no significant 
trend of the mean offset with \logg. From these findings, we decided to adjust our estimated LR LTE \nafe\ 
by 0.16 dex to place it on the same abundance scale as for the HR LTE APOGEE sample.

To compare with the GALAH data, after the 0.16 dex offset described above is applied to our 
LR LTE estimates, we corrected them to the NLTE scale following the recipe described in 
Section \ref{nlte}. The Na abundance differences with respect to GALAH and their 
trend with \logg\ are displayed in the right column of panels shown in Figure \ref{nafehr}.  
Our LR NLTE Na abundances exhibit a very small systematic offset of 0.04 dex, as shown  
in the top panel. The lower panel, however, indicates a negative trend with \logg. 
In order to eliminate this trend, we used a linear fit, with slope and intercept of 
--0.21 dex dex$^{-1}$ and 0.88 dex, respectively, and applied it to our LR NLTE values.
We noticed no significant trend with \teff\ or \feh\ after this correction was applied.

To summarize our approach for determination of the final adopted \nafe\ estimates, we first add 0.16 dex to our 
measured LR \nafe, obtain the NLTE-corrected value, and apply the correction function found in 
the bottom-right panel of Figure \ref{nafehr} to determine the final LR NLTE \nafe\ estimate. 
We use this final value for the following analysis. The comparisons with 
the high-resolution data and the other two tests described in Sections \ref{syserror} and 
\ref{snrtest} suggest that the expected precision of the estimated [Na/Fe] is 
better than 0.3 dex for [Fe/H] $>$ --2.0 and SNR $>$ 10.

\subsection{Accuracy of the Stellar Parameters Determined by the SSPP}

We also compared the stellar parameters delivered by the SSPP 
to those from the APOGEE and GALAH surveys to check their accuracy and 
uncertainties, because any systematic 
error in the SSPP stellar parameters can result in systematic deviations 
in the determined [Na/Fe], as demonstrated in Section \ref{syserror}. 
Note that we employed the SDSS stars in common with stars from 
the APOGEE and GALAH surveys with [Na/Fe] available, as we used them to calibrate our estimates of 
\nafe. Figure \ref{tgm} shows the distributions of the differences in 
\teff\ (left), \logg\ (middle), and \feh\ (right) for 
the APOGEE (upper panels) and GALAH (lower panels) data. The mean difference ($\mu$) 
and the standard deviation ($\sigma$) are derived from a Gaussian fit to each 
distribution. We note the very small mean offsets and scatters for both 
surveys. Specifically, the temperature comparison indicates that the zero-point 
offset is less than 30 K, with a scatter less than 90 K. The gravity comparison 
shows $\mu <$ 0.1 dex with $\sigma <$ 0.25 dex, and  
$\mu <$ 0.05 dex with $\sigma <$ 0.15 dex for the metallicity. 
Note that the metallicity from the GALAH survey is the NLTE value. As the NLTE 
correction for [Fe/H] is very small (mostly less than 0.1 dex) for the stars 
with [Fe/H] $>$ --1.0 \citep{lind2012} and most of the GALAH stars have [Fe/H] $>$ --1.0, 
we simply used the NLTE value of [Fe/H] for this comparison.

Taking into account that the SDSS stellar spectrum is low resolution, these 
mean offsets and scatters are quite small, suggesting that the SSPP recovers estimates 
of the stellar parameters very well.

\begin{figure}[!t]
\centering
\includegraphics[width=1.0\linewidth]{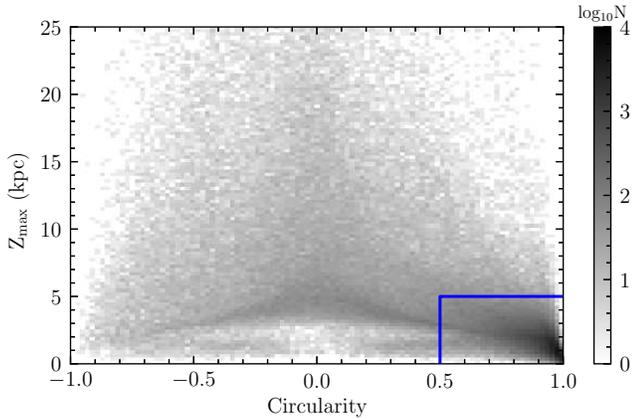}
\caption{Density map of our sample stars in $Z_{\rm{max}}$ versus circularity space.
Positive and negative circularity values correspond to prograde and retrograde motions, respectively.
The stars inside the blue box are regarded to be of $in\,situ$ origin, while the
objects outside the box are taken to be of accreted origin.
Note that, although a number of stars with $r_{\rm{max}} < 3.5$ kpc are located outside the blue box, 
they are regarded as belonging to the $in~situ$ origin (see text).}
\label{sample}
\end{figure}

\begin{figure*}
\centering
\includegraphics[width=0.9\linewidth]{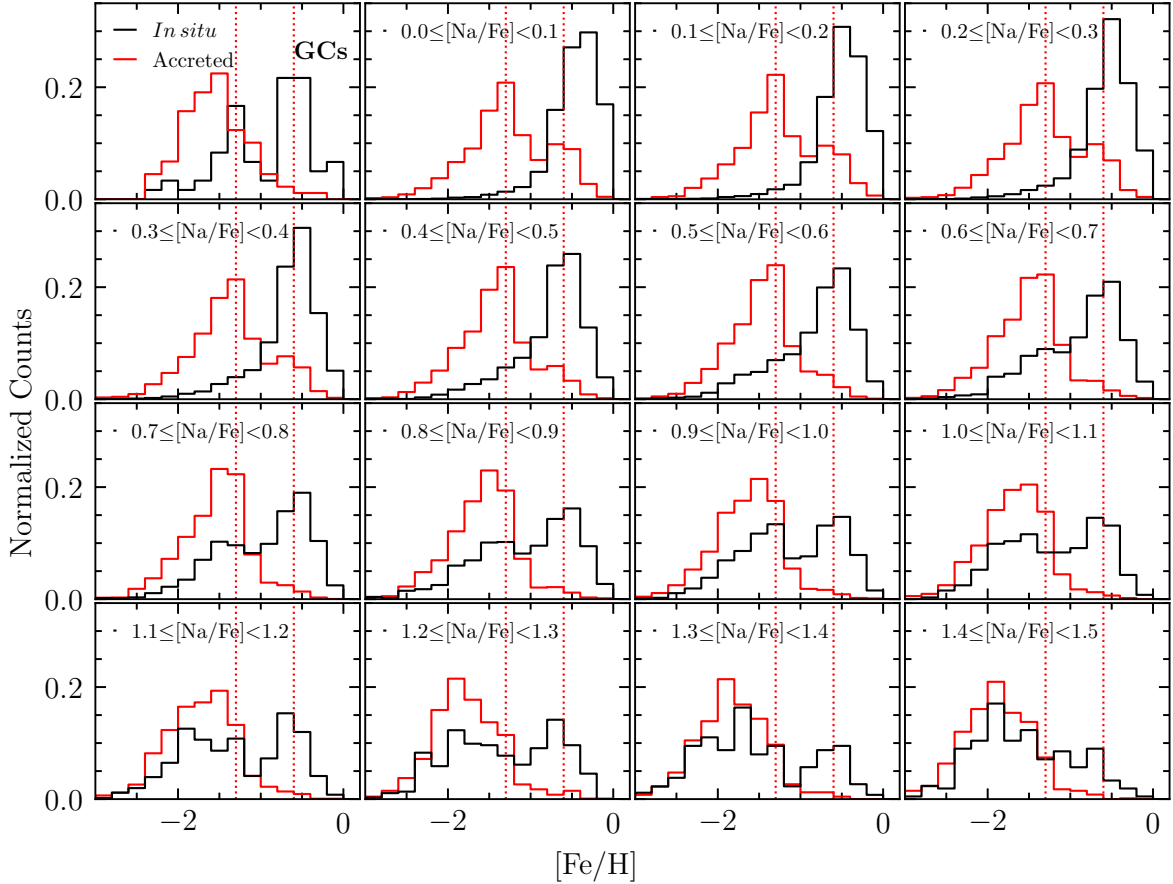}
\caption{Metallicity distribution functions (MDFs) of the $in\,situ$ (black) and the accreted (red)
components, which are grouped by the dividing scheme of \citet{massari2019}.
The histogram is normalized by the total number of stars in each group.
The top-left panel shows the MDFs of the GCs. The $in~situ$ GCs display two
peaks around [Fe/H] = --1.3 and --0.6, respectively, whereas the accreted component exhibits 
one peak at [Fe/H] = --1.6. The two vertical dotted reference lines are provided at [Fe/H] = --1.3 and --0.6, 
respectively. The other panels present the MDFs of our sample of stars in different ranges of [Na/Fe]. 
Notice the rise of the second peak in the $in~situ$ sample from [Na/Fe] = +0.6; as the Na-abundance 
ratio increases, the portion of the metal-poor stars increases as well. In comparison, 
the accreted population displays two peaks at low [Na/Fe], and they 
disappear above [Na/Fe] $\sim$ +0.5; its MDF is gradually shifted 
to the low-metallicity regime with increasing [Na/Fe].}
\label{mdf2g}
\end{figure*}

\section{An Example Application}\label{sec4}

Here we describe an example application of our sodium-abundance ratio estimates for a large sample of stars with 
available low-resolution spectra from SDSS. Other applications are clearly of interest, which we will 
pursue in future papers. When the currently observed GCs in the MW are divided into the \insitu\ and accreted component by 
their dynamical properties \citep[e.g.,][]{massari2019}, their MDFs are dissimilar, with 
distinct characteristics. As the Na-rich stars are believed to have originated from stripped or totally disrupted GCs, 
we might expect that their dynamical properties and chemical characteristics may be 
similar to those observed in the surviving GCs, but different for the Na-normal stars. We explore this hypothesis below.

\subsection{The Field Star Sample}

We applied our method for determining [Na/Fe] to 539,920 stellar spectra with SNR $\geq$ 10, which were 
assembled from the SDSS stars. We adopted parallaxes 
from $Gaia$ EDR3 to calculate the distances, which 
we require to estimate the strength of the ISS lines 
from the 3D reddening map. When calculating the distance, we corrected 
the global parallax offset of $+$0.017 mas \citep{lindegren2021}.
After the corrections for NLTE effects and the 
systematic offset described in Sections \ref{nlte} and \ref{hrcheck}, respectively, 
we are left with a total number of 427,799 stars with available LR NLTE \nafe\ values.

\subsection{Calculation of Space Velocities and Orbital Parameters}

For the dynamical study of our program stars 
we require information on their velocity components and orbital parameters.
We employed the proper motions and parallaxes from $Gaia$ EDR3 
and the radial velocities from the SDSS spectra, and computed 
the $U, V, W$ velocity components and dynamical parameters, including 
the maximum (\rmax) and minimum (\rmin) distances from 
the Galactic center, maximum distance (\zmax) from the Galactic plane, 
orbital eccentricity ($e$), angular momentum, and energy, etc., 
using the galpy Python package\footnote{\url{http://github.com/jobovy/galpy}} \citep{bovy2015}, 
adopting the Galactic gravitational potential model of \citet{mcmillan2017}.
During these calculations, we integrated the orbit of each star 
over 13 Gyr with 1 Myr steps, and assumed $V_{\rm{LSR}}$ = 233.1 \kms\ 
for the Local Standard of Rest, a solar peculiar 
motion of ($U,V,W$)$_\odot$ = (11.1, 12.24, 7.25) km\,s$^{-1}$ \citep{schonrich2010}, 
and the distance of the Sun from the Galactic center of $R_\odot$ = 8.21 kpc. The 
orbital parameters of Galactic GCs were also computed by adopting 
their positions, proper motions, distances, and radial velocities from 
\citet{vasiliev2019}.


\begin{figure*}
\centering
\includegraphics[width=0.8\linewidth]{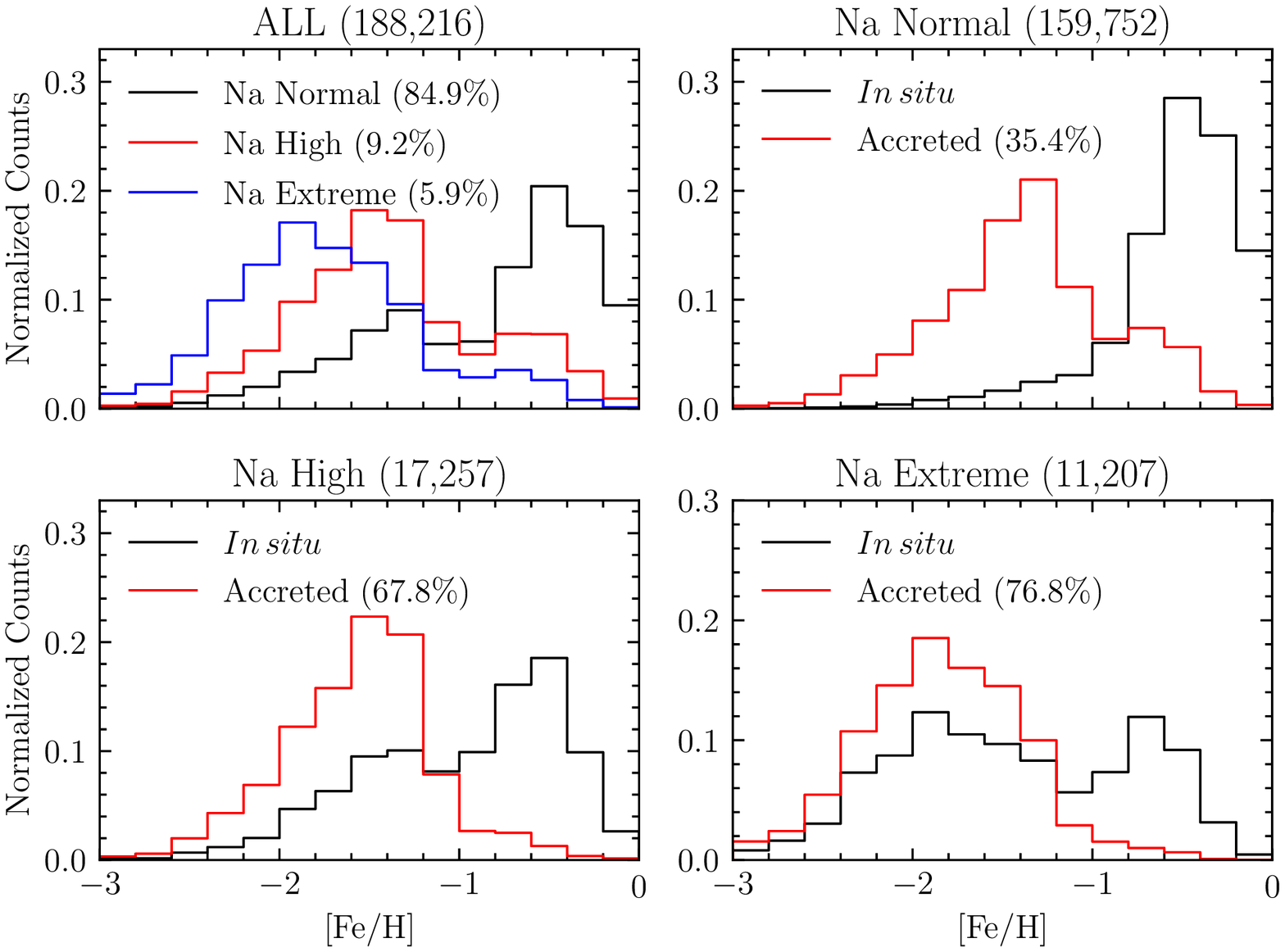}
\caption{Metallicity distribution functions (MDFs) of our sample according to their level of 
Na enhancement: all (upper left), [Na/Fe] $<$ +0.6 (upper right), +0.6 $\leq$ [Na/Fe] $<$ +1.0 (lower left), 
and [Na/Fe] $\geq$ +1.0 (lower right). It is clear to see in the upper-left 
panel that all three subgroups have different peaks and shapes of their MDFs. About 85\,\% of 
our sample stars belong to the Na-normal subgroup, and only 15\,\% of stars have [Na/Fe] $\geq 
+0.6$. The other panels show the MDFs of the $in\,situ$ and accreted stars for the three 
subgroups. The fraction of the accreted stars clearly increases with increasing [Na/Fe].}
\label{mdf3g}
\end{figure*}


\subsection{Results and Discussion}

Among the SDSS stars with estimated [Na/Fe] and available orbital parameters, 
we selected the stars with 4400 K $\leq$ $T_{\rm{eff}}$ $\leq$ 7000 K, 
--3 $\leq$ [Fe/H] $\leq 0$, and SNR $\geq$ 25.0 to ensure 
the most reliable measurement of \nafe. We also imposed 
on our sample stars the requirement of the renormalized unit weight error less than 1.4 from $Gaia$ EDR3 
to select stars with good astrometric solutions. For stars with derived 
\rmax\ $\leq$ 200 kpc, we ended up with a final sample of 188,216 stars.

Provided that the Na-rich stars originated from GCs or disrupted GCs, we may expect that 
their MDF and dynamical properties would be similar to those of the GCs. In addition, 
they would exhibit distinct features in their MDF, compared to those from the Na-normal 
stars. To identify their connection to the GCs, we followed a similar approach 
as \citet{massari2019}. They divided the GCs into $in\,situ$ (bulge and disk clusters) and accreted 
clusters, using their dynamical properties. They defined that the bulge clusters have small 
orbits with \rmax\ $< 3.5$ kpc, and disk clusters have orbital circularity larger than 0.5 and 
$Z_{\rm{max}} < 5$ kpc. Note that the orbital circularity is defined 
as $L_Z/L_{Z, \rm{circ}}$, where $L_{Z, \rm{circ}}$ is the angular momentum of 
a circular orbit with a cluster's energy. 
Figure \ref{sample} displays the density map of our sample stars on $Z_{\rm{max}}$ 
versus circularity space. In the figure, positive and negative 
circularity values mean prograde and retrograde orbits, respectively. The 
stars inside the blue box are classified as of the $in\,situ$ origin, while outside 
the box they are considered of accretion origin.


After dividing into the two groups, we first inspected the MDFs of each group in different 
bins of [Na/Fe] values, as shown in Figure \ref{mdf2g}. The top-left panel of the figure 
shows the MDFs of the GCs in the MW. The black histogram is for the $in\,situ$ GCs, 
while the red histogram is for the accreted GCs, divided by following the work by \citet{massari2019}. 
The [Fe/H] values of the GCs were adopted from the \citet{harris1996,harris2010} catalog. 
The notable features in this panel are that the MDF of the accreted GCs has only a single peak 
at [Fe/H] $\sim$ --1.6, with more objects populating the more metal-poor 
region, while the MDF of the $in\,situ$ GCs exhibits two peaks at 
[Fe/H] = --1.3 and --0.6, respectively, indicated by the two 
vertical dotted lines. The rest of the panels in the figure 
show the MDFs of our program stars.

Inspection of the MDFs in each panel reveals that the MDF for the \insitu\ 
population of stars peaks at between [Fe/H] = --0.6  and --0.4, with 
a long tail toward the low-metallicity region, up to [Na/Fe] = +0.6. These Na-normal, metal-rich 
stars are the dominant sources of the canonical Thick Disk (TD) and Metal-Weak 
Thick Disk \citep[MWTD;][]{carollo2019}. The second peak at [Fe/H] = --1.3 starts to rise from [Na/Fe] = +0.6, and 
the first peak remains the same, at [Fe/H] = --0.6, for \nafe\ $>$ +0.6. 
These features qualitatively agree with those of the $in~situ$ GCs, as can 
be seen in the top-left panel. From [Na/Fe] = +1.0 to \nafe\ = +1.5, the second peak becomes more 
obvious, with a much broader, but more metal-poor, distribution, while the fist peak still exists.
The much larger scatter and more metal-poor feature in the second peak may imply that more than one 
population contributes to the second peak. Even though they are defined as the $in~situ$ 
component, it is possible that some of the metal-poor component came from the 
accreted GCs, as claimed by \citet{woody2021}, who argue that the metal-poor GCs 
are mostly accreted from disrupted dwarf galaxies.

The accreted population of our sample fairly well follows the shape 
of the MDF of the accreted GCs above [Na/Fe] = +0.5. For \nafe\ $<$ +0.5, 
we can also observe a second peak at [Fe/H] = --0.6. Considering their metallicity and 
distinct dynamical status, this group of stars may consist of the GE \citep{helmi2018} and 
Splashed Disk \citep[SD;][]{belokurov2020}, and the rest of the more metal-poor stars may contribute 
to the Sequioa \citep{myeong2019} and Thamnos 1 \& 2 \citep{koppelman2019} structures.  We note 
in Figure \ref{mdf2g} that the low-metallicty tails of the MDFs of the $in~situ$ 
and the accreted populations in the range of [Na/Fe] $>$ +0.9 exhibit similar peaks and shapes. 
This analogy may suggest that the low-metallicity, Na-rich stars may share chemically similar 
GC progenitors. 

For the ensemble view of the $in~situ$ and the accreted populations of our sample, 
based on the MDF characteristics found in Figure \ref{mdf2g}, we  
divided our sample into three subgroups: Na-normal stars 
with [Na/Fe]$<$ +0.6, Na-high stars with +0.6 $\leq$ [Na/Fe] $<$ +1.0, 
and Na-extreme stars with [Na/Fe] $\geq$ +1.0, and explored differences 
in their MDFs, as shown in Figure \ref{mdf3g}.

The upper-left panel of Figure \ref{mdf3g} displays the MDFs of the three subgroups, 
normalized by the total number of stars in each subgroup. Different peaks and 
morphologies for the MDFs are evident. The general behavior is that the 
higher the sodium-abundance ratio, the lower the metallicity peak. About 
85\% of our sample stars belong to the Na-normal subgroup, and only 15\% of stars 
have [Na/Fe] $\geq$ +0.6. The other three panels exhibit the MDFs of the $in\,situ$ 
(black) and accreted (red) stars for each subgroup. The $in\,situ$ stars of the Na-normal 
subgroup (upper right) exhibits a single peak at [Fe/H] $\sim$ --0.5 with a long metal-poor wing. 
This group of stars comprises mainly the canonical TD and MWTD.
On the other hand, the accreted subgroup displays a single peak at [Fe/H] $\sim$ --1.3, without 
much skewness. This subgroup comprises the GS, GE, and SD stars. 
The two MDFs are reminiscent of the local disk and halo populations in the MW. 

About 9.2\% of stars (lower-left panel of Figure \ref{mdf3g}) among our sample belong to the 
Na-high subgroup. The shapes of the MDFs of the $in\,situ$ and accreted stars in this subgroup 
are very similar to those of the GCs (see the upper-left panel of Figure \ref{mdf2g}) --- 
a single peak around [Fe/H] = --1.6 for the accreted population and the double peaks around 
[Fe/H] = --1.3 and --0.6 for the $in\,situ$ population. As Galactic field stars mostly have 
[Na/Fe] $<$ +0.5 \citep{suda2008}, and the Na-high stars have the same characteristics 
in their MDF as the Galactic GCs, we can confirm that the stars in the Na-high subgroup are likely
the escaped stars from present GCs.

The MDFs of the Na-extreme subgroup (lower-right panel of Figure \ref{mdf3g}) present some distinct features 
compared to the Na-normal and Na-high subgroups. Although the $in~situ$ population still 
has two peaks, the metal-poor stars are more prominent. The accreted 
component is also more metal-poor and has a broader distribution. 
One notable feature is that the metal-poor component ([Fe/H] $<$ --1.2) 
of the MDF for the $in~situ$ population generally follows that of the accreted component, 
a characteristic that does not appear for the surviving GCs. Consequently, it is plausible 
that the metal-poor stars in the Na-extreme subgroup may have originated from 
already disrupted metal-poor GCs, and experienced similar chemical evolution, 
but different dynamical evolution. Another possible interpretation is 
that the $in~situ$ metal-poor stars may not be the stars formed 
in \insitu\ GCs. According to the analysis of 
\citet{piatti2019}, among the prograde GCs, which are assumed to be of 
the $in~situ$ origin, the number of accreted GCs is similar to that 
of the $in~situ$ ones. Following this reasoning, some of the stars in our metal-poor, 
Na-extreme $in~situ$ population may originate from accreted GCs. 
This interpretation is in line with the claim by \citet{woody2021}.

Regarding the metal-rich ([Fe/H] $>$ --1.0) \insitu\ population 
in the Na-high and Na-extreme subgroups, we note that these may 
be related to the metal-rich ([Fe/H] $>$ --0.7) population recently discovered 
from APOGEE data by \citet{trincado2021}. These authors claimed that this 
population is likely to be stellar debris from partially or fully disrupted metal-rich GCs.

Another point to consider is that some fraction 
of Carbon-Enhanced Metal-Poor \citep[CEMP;][]{beers2005} stars 
exhibit enhancements of sodium although many of the CEMP stars exhibit 
[Na/Fe] $<$ 0.5 \citep{spite2014,hansen2015,yoon2016,purandardas2021}.
Thus, our Na-extreme stars may also be related to CEMP stars. 
Indeed, when we considered as CEMP stars the objects with [Fe/H] $<$ --1.0 and [C/Fe] $>$ +0.7, 
we found from the Na-extreme subgroup that the fraction of CEMP stars for 
[Fe/H] $<$ --1.0 is 12.0\%; it increases 
to 18.7\% for [Fe/H] $<$ --2.0. This proportion is much higher than those of previous 
studies \citep[e.g.,][]{lee2013}. We also found that the CEMP fraction (19.1\%) of the 
accreted component in the Na-extreme subgroup is higher than that (16.3\%) of the $in~situ$ component 
for [Fe/H] $<$ --2.0. In the metallicity range of [Fe/H] $>$ --2.5, the CEMP-$s$ (those 
enhanced with slow neutron-capture elements) stars are dominant among the different CEMP 
subclasses (\citealt{aoki2007}; see also \citealt{yoon2016,yoon2018}); hence their 
likely progenitors are asymptotic giant branch (AGB) stars. 
Furthermore, the second-generation stars observed in GCs exhibit low carbon and high nitrogen abundances.
These results suggest that many of the Na-extreme stars probably did not originate from GCs. 

Taking into account only the Na-high stars, we obtained the 
fraction of 9.2\% (17257/188216), which is within the fractions 
reported by previous studies \citep{carretta2010, martell2011,ramirez2012}. This fraction implies 
a non-negligible contribution from the GCs to the build-up of the Galactic halo.

To sum up, since the $in~situ$ and accreted stars of our Na-high stars exhibit 
similar MDFs to those of present GCs, they may have originated from these, but this is not 
the case for the Na-extreme stars. As the fraction of the Na-high stars is 9.2\%, 
the contribution from GC-origin stars to the build-up of the 
Galactic halo appears non-negligible. Furthermore, because the metal-poor stars in 
the $in~situ$ and accreted populations exhibit similar 
MDFs, we cannot rule out that they may share the progenitors with similar chemical-evolution histories. 
Indeed, a K-S two-sample test between MDFs for  
the $in~situ$ and accreted stars with [Fe/H] $<$ --1.2 in the 
Na-extreme subgroup yielded a $p$-value of 0.0722, which is small, but only 
permits a marginal rejection of their selection from the same parent population.

\section{Summary}\label{sec5}

We have presented a method for the estimation of [Na/Fe] from 
low-resolution stellar spectra. To determine [Na/Fe] more accurately, we developed 
a technique of removing the ISS lines in a 
stellar spectrum, using the correlation between the Galactic reddening, $E(B-V)$, 
and the strength of \nai\ D doublet lines produced by the ISM. In addition, 
we applied an NLTE correction to determine [Na/Fe] more accurately and precisely.
We also investigated the MDFs of the Na-rich stars, and compare with those of 
Galactic GCs, in order to better understand the connection of the Na-rich stars with 
the buildup of the Galactic halo. In a subsequent paper, we will carry out a
more thorough analysis of the Na-rich stars gathered from SDSS and LAMOST, using their 
chemical and dynamical properties to unravel their relation to the GCs and 
the Galactic halo assembly.

We also note that our measured [Na/Fe] will be extremely 
useful to calibrate photometrically estimated [Na/Fe] for stars 
from narrow-band photometric surveys such as 
Javalambre Photometric Local Universe Survey \citep[J-PLUS;][]{cenarro2019} 
and the Southern Photometric Local Universe Survey \citep[S-PLUS;][]{mendes2019}. 
Eventually, these efforts should result in tens to hundreds of million stars 
with measured [Na/Fe] to explore various stellar populations in the MW.

\begin{acknowledgements}

Y.S.L. acknowledges support from the National Research Foundation (NRF) of
Korea grant funded by the Ministry of Science and ICT (NRF-2021R1A2C1008679). 
J.-R.K. acknowledges support from Basic Science Research Program through 
the National Research Foundation of Korea(NRF) funded by the Ministry of 
Education (NRF-2019R1I1A3A02062242). T.C.B. acknowledges partial support for this work
from grant PHY 14-30152; Physics Frontier Center/JINA Center for the Evolution
of the Elements (JINA-CEE), awarded by the U.S. National Science Foundation.

This work has made use of data from the European Space Agency (ESA) mission
{\it Gaia} (\url{https://www.cosmos.esa.int/gaia}), processed by the {\it Gaia}
Data Processing and Analysis Consortium (DPAC, \url{https://www.cosmos.esa.int/web/gaia/dpac/consortium}). 
Funding for the DPAC has been provided by national institutions, in particular the institutions
participating in the {\it Gaia} Multilateral Agreement.

Funding for the Sloan Digital Sky Survey IV has been provided by the 
Alfred P. Sloan Foundation, the U.S. Department of Energy Office of 
Science, and the Participating Institutions. 

SDSS-IV acknowledges support and resources from the Center for High 
Performance Computing  at the University of Utah. The SDSS 
website is www.sdss.org.

SDSS-IV is managed by the Astrophysical Research Consortium 
for the Participating Institutions of the SDSS Collaboration including 
the Brazilian Participation Group, the Carnegie Institution for Science, 
Carnegie Mellon University, Center for Astrophysics | Harvard \& 
Smithsonian, the Chilean Participation Group, the French Participation Group, 
Instituto de Astrof\'isica de Canarias, The Johns Hopkins 
University, Kavli Institute for the Physics and Mathematics of the 
Universe (IPMU) / University of Tokyo, the Korean Participation Group, 
Lawrence Berkeley National Laboratory, Leibniz Institut f\"ur Astrophysik 
Potsdam (AIP),  Max-Planck-Institut f\"ur Astronomie (MPIA Heidelberg), 
Max-Planck-Institut f\"ur Astrophysik (MPA Garching), 
Max-Planck-Institut f\"ur Extraterrestrische Physik (MPE), 
National Astronomical Observatories of China, New Mexico State University, 
New York University, University of Notre Dame, Observat\'ario 
Nacional / MCTI, The Ohio State University, Pennsylvania State 
University, Shanghai Astronomical Observatory, United 
Kingdom Participation Group, Universidad Nacional Aut\'onoma 
de M\'exico, University of Arizona, University of Colorado Boulder, 
University of Oxford, University of Portsmouth, University of Utah, 
University of Virginia, University of Washington, University of 
Wisconsin, Vanderbilt University, and Yale University.
\end{acknowledgements}

\newpage

 \end{document}